\documentclass[amsmath,amssymb,aps,prc,twocolumn,superscriptaddress,10pt]{revtex4-1}

\usepackage{graphicx}
\usepackage{dcolumn}
\usepackage{bm}
\usepackage{enumitem}
\usepackage{textcomp}
\usepackage{gensymb}
\usepackage{appendix}
\usepackage{soul}
\usepackage[normalem]{ulem}

\newcommand{\lpc}{Universit\'e de Caen Normandie, ENSICAEN, CNRS/IN2P3, LPC Caen,
UMR6534, 14000 Caen, France}
\newcommand{\cea}{Universit{\'e} Paris-Saclay, CEA, List, Laboratoire National Henri Becquerel (LNE-LNHB), 91120 Palaiseau, France}
\newcommand{\msu}{Facility for Rare Isotope Beams and Department of Physics
and Astronomy, Michigan State University, East Lansing 48824 MI, USA}
\newcommand{\ganil}{GANIL, CEA/DRF-CNRS/IN2P3, 14076 Caen, France}
\newcommand{\leuven}{KU Leuven, Instituut voor Kern- en Stralingsfysica, 3001 Leuven, Belgium}
\newcommand{\irlnpa}{International Research Laboratory Nuclear Physics and Nuclear Astrophysics, CNRS-MSU, East Lansing 48824 MI, USA}
\newcommand{\nist}{Radiation Physics Division, National Institute of Standards and Technology, 100 Bureau Drive, Gaithersburg, Maryland 20899, USA}

\usepackage{xcolor}
\usepackage{amsmath}
\usepackage{physics}
\usepackage{bm}

\begin{document}

\preprint{backscattering_pp}

\title{Backscattering Study of Electrons from 0.1 to 3.4 MeV}

\affiliation{\lpc}
\affiliation{\msu}
\affiliation{\irlnpa}
\affiliation{\nist}
\affiliation{\cea}
\affiliation{\ganil}
\affiliation{\leuven}

\author{M.~Kanafani}
\affiliation{\lpc}
\author{X.~Fl\'echard}
\email{Corresponding author: flechard@lpccaen.in2p3.fr}
\affiliation{\lpc}
\author{O.~Naviliat-Cuncic}
\affiliation{\lpc}\affiliation{\msu}\affiliation{\irlnpa}
\author{R.~Garreau}
\affiliation{\lpc}
\author{T.E.~Haugen}
\affiliation{\msu}\affiliation{\nist}
\author{L.~Hayen}
\affiliation{\lpc}
\author{S.~Leblond}
\affiliation{\cea}
\author{E.~Li\'enard}
\affiliation{\lpc}
\author{X.~Mougeot}
\affiliation{\cea}
\author{G.~Qu\'em\'ener}
\affiliation{\lpc}
\author{A.~Rani}
\affiliation{\lpc}
\author{J-C.~Thomas}
\affiliation{\ganil}
\author{S.~Vanlangendonck}
\affiliation{\leuven}

\date{\today}
\begin{abstract}
Benchmarking
simulation codes for electron transport and scattering
in matter is a crucial step for estimating uncertainties in many applications.
However, experimental data for electron energies of a few MeV
is scarce to make such comparisons.
We report here the measurement and the quantitative analysis
of backscattering probabilities
of electrons in the energy range 0.1 to 3.4~MeV impinging on a
YAP:Ce scintillator. The setup consists
of a $2\times 2\pi$ calorimeter which enables, in particular, the inclusion
of large incidence angles. The results are used to benchmark various
scattering models incorporated in Geant4, showing relative deviations smaller
than 5\% between experiment and simulations. They demonstrate the current
rather high reliability
of the simulations when employing appropriate electromagnetic Physics Lists.
\end{abstract}
\maketitle
\section{Introduction}

Tracking particles through matter using Monte Carlo transport codes is
an essential tool for designing experiments and for analyzing data in various fields,
including high-energy physics, astroparticle physics, nuclear physics,
as well as space and medical applications. Among the many physical processes
involved in the simulations, electron scattering in matter is a critical component.
Benchmarking the performance of scattering models in simulation codes is
crucial for estimating uncertainties associated with the results from such codes.
For example, precision correlation measurements in nuclear and neutron beta decay
\cite{Fle11, Fen18, Ara20, Sau20, Mue22, Bur22, Sun22} are often limited by the estimated
uncertainties associated with simulations. These
must accurately model the scattering of electrons or positrons in the detectors and
surrounding materials. Considering the wide range of beta transitions of interest,
obtaining benchmarking data for energies in the few MeV range is important.

Several dedicated electron backscattering studies have been carried out in
the past \cite{Tab67, Mar03, Mar06, Sot13, Spr24}, with various degrees of quantitative analyses.
Wright and Trump reported measurements on thick metal targets, at normal
incidences and for five energies between 1 and 3~MeV \cite{Wri62}, but no
quantitative comparison was provided. Such measurements were extended by Tabata
from 3.4 to 14~MeV on various targets \cite{Tab67}, indicating significant
deviations from simple calculations, in particular for the angular distributions.
Martin {\em et al.} performed measurements at the single energy of 124~keV,
in beryllium and silicon \cite{Mar03} and in plastic scintillators \cite{Mar06}
at normal incidence. Due to significant
systematic uncertainties (up to 23$\%$), the comparison to Geant4 and Penelope
simulations
used free scaling factors, which makes it difficult to provide a precise
quantitative assessment of the overall agreement.
Relative comparisons with Geant4 simulations have also been reported in
Ref.~\cite{Sot13} from measurements using $^{60}$Co and $^{207}$Bi sources on
silicon detectors. Because of the complexity of the source spectra,
the conclusions are not straightforward
and no quantitative benchmark on backscattering is provided.
A dedicated spectrometer has been designed for this purpose~\cite{Loj15} but
no systematic quantitative study of backscattering has been published so far
with such a tool.
More recently, Spreng {\em et al.} performed measurements  using an electron gun, finding consistent results with Geant4 simulations within the quoted uncertainties \cite{Spr24}. However, this study covered electron energies up to 10~keV. A comprehensive open-access database, compiling electron backscattering coefficient measurements published before 2023 is available in Ref. \cite{Akb23}.

Several systematic evaluations of the Geant4 performance, using different electromagnetic
physics models and for electron energies extending up to 15~MeV, have been published in the
last decade \cite{Kim15, Bas16, Don18}. These stressed in particular how critical some geometry and tracking parameters could be, for several
versions of Geant4. The review by Dondero {\em et al.} showed that
measurements conducted in the 1960s and the 1970s, display
variations of up to 22\% in the backscattering probabilities, 
for energies smaller than 100~keV at normal incidence on aluminum targets \cite{Don18}.
Some of these measurements are
fairly well reproduced by Geant4 simulations. 
The most stringent tests, in the energy range from 100~keV to 1~MeV
\cite{Don18}, used measurements carried out at the Sandia National Laboratories
\cite{Loc80, Loc81} on several targets (C, Al, Mo, Ta and U) at normal incidence,
as well as for incident angles up to $75^\circ$ for Al targets. Comparison with
the Geant4 single scattering model at normal incidence shows generally good
agreement \cite{Don18}, with relative deviations mostly below $5\%$ for the
low-mass targets (C and Al). However, discrepancies up to $15\%$ were observed
for higher mass targets or different incidence angles, in particular for
energies below 200~keV. 

In the present work, we have studied the backscattering
of electrons in the energy range 0.1 to 3.4 MeV, including incidence angles
from 0 to $90^\circ$. The measurements have been carried out using
beta particles from $^6$He decay and
YAlO$_3$ Ce-doped inorganic scintillators (YAP) \cite{Mos98,Men98} which
serve as active scattering media.
The results have been compared with predictions from
different electromagnetic Physics Lists of Geant4, to benchmark
this code in the considered energy range.
The experimental data that support the findings of this article are openly  available \cite{dataGANIL}.
 
\section{Experimental setup}

The experimental setup is described in detail in Ref.~\cite{Kan22}.
Only key features are summarized below.

A 25 keV $^6$He$^+$ ion beam,
with a typical intensity of $10^4$~pps, was produced and
delivered
by the LIRAT-SPIRAL1 beam line at the GANIL facility in Caen, France.
The beam was chopped
with a fast electrostatic deflector, enabling the implementation of
cycles with implantation and decay phases.

\begin{figure}[!htb]
\includegraphics[width = 1.\columnwidth]{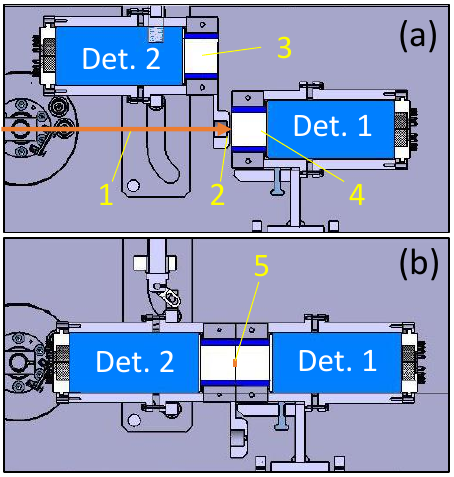}
\caption{\label{fig:setup}
Sectional views of the detection setup for respectively (a) the
implantation and, (b) the detection configurations. The labels indicate the
$^6$He$^+$ beam (1), the \O~4~mm collimator (2), the two YAP scintillators
(3,4) and the beam implantation region (5).}
\end{figure}

Before entering the detection chamber (Fig.\ref{fig:setup}),
the beam
was collimated using two sets of \O~6~mm rings. It was further
collimated
by a \O~4~mm aperture and implanted on the surface of a YAP crystal
scintillator during a time interval of 2.5~s.

The implantation detector (Det.~1 in Fig.~\ref{fig:setup}) was fixed in
the chamber.
After implantation, a movable detector (Det.~2),
mounted on a fast actuator,
was brought into contact with the fixed detector [Fig.\ref{fig:setup} (b)],
creating a 4$\pi$ geometry around the location of implantation.
Moving Det.~2 from the implantation to the detection configurations
took about 0.8~s, and returning to the implantation configuration
took 1.7~s. The total duration of the
cycle was 17~s. In the analysis, a decay time window of 12~s was selected
during the beam-off phase.

\begin{figure}[!htb]
\includegraphics[width = 1.\columnwidth]{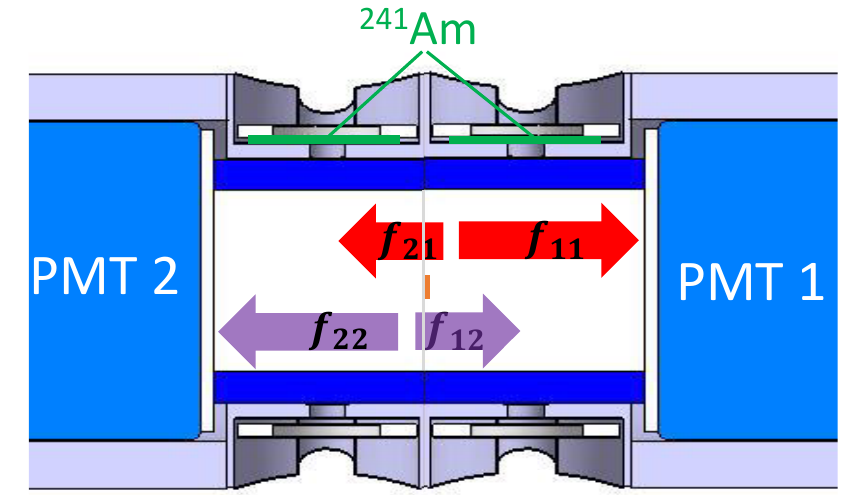}
\caption{\label{fig:Det}
Sectional view of the detector assembly in the measuring
configuration. The red and purple arrows indicate the scintillation
light distributed between the two detectors. The $f_{ij}$ coefficient inside
an arrow corresponds to the fraction of light collected by PMT $i$
and originating from YAP crystal $j$.}
\end{figure}

Each detector consists of a cylindrical (\O30~mm $\times$ 30~mm) YAP:Ce scintillator,
surrounded by a plastic (PVT) scintillator (EJ-204), which serves as a veto
(Fig.~\ref{fig:Det}). Both scintillators are arranged in a phoswich configuration
and are readout by a single Hamamatsu R7723 photomultiplier tube (PMT).
The signals from the PMTs, along with a fast signal synchronized with the
chopper, were sent to three channels of the FASTER 
digital data acquisition system \cite{FASTER} configured in a TDC/QDC mode.
A pulse shape analysis technique \cite{Kan22} was implemented to discriminate signals
originating from the YAP or from the PVT, based on their different time responses
(27~ns decay time for YAP and 2.5~ns for PVT).
Two 5~kBq $^{241}$Am calibration sources, fixed on the side of each detector,
were used to perform crude energy calibrations and to correct for gain
variations. No light reflector was used on the contact surface of the scintillators,
ensuring that no dead layer affects the electron energy measurements.

According to TRIM calculations \cite{TRIM}, the 25~keV $^6$He$^+$ beam was
implanted at
a depth of 130~nm inside Det.~1. The diffusion of $^6$He atoms was estimated to
be negligible \cite{Kan22}.
The decay of $^6$He nuclei provided a source of electrons
with kinetic energies up to 3.5~MeV.
The two YAP scintillators, which enclose the $^6$He source,
serve both as scattering targets and as calorimeters.
The backscattering probability is deduced from events recorded in coincidence
between the two YAP,
for which the electrons deposited some energy in each detector. 
Since beta particles are emitted isotropically over a 4$\pi$~sr solid angle, the
backscattering probabilities measured as a function of the total recorded energy
encompass also large incidence angles (up to $90^\circ$), which have not been
accessed experimentally so far at these energies.

\section{Experimental Data}

Due to the optical coupling between the two
scintillators, a fraction ($\approx 20\%$)
of the light emitted by one scintillator crosses the interface and
is collected by the PMT of the opposite detector. This is schematically shown
by the red and purple arrows in Fig.~\ref{fig:Det}. Coincidence
events show that both detection channels are triggered provided the
total energy deposited is larger than 50~keV.

The construction of the main 2D histogram, used in the data analysis,
involves three
main steps: the background subtraction, the detector calibration and the
crosstalk correction. However, as shown in Appendix \ref{app:crosstalkAndResolution},
the last two steps are coupled and must be addressed simultaneously.
The background sources and the procedure for background subtraction are detailed
in Appendix \ref{app:bckgd_sub}.

\begin{figure}[!htb]
\includegraphics[width = 1.\columnwidth]{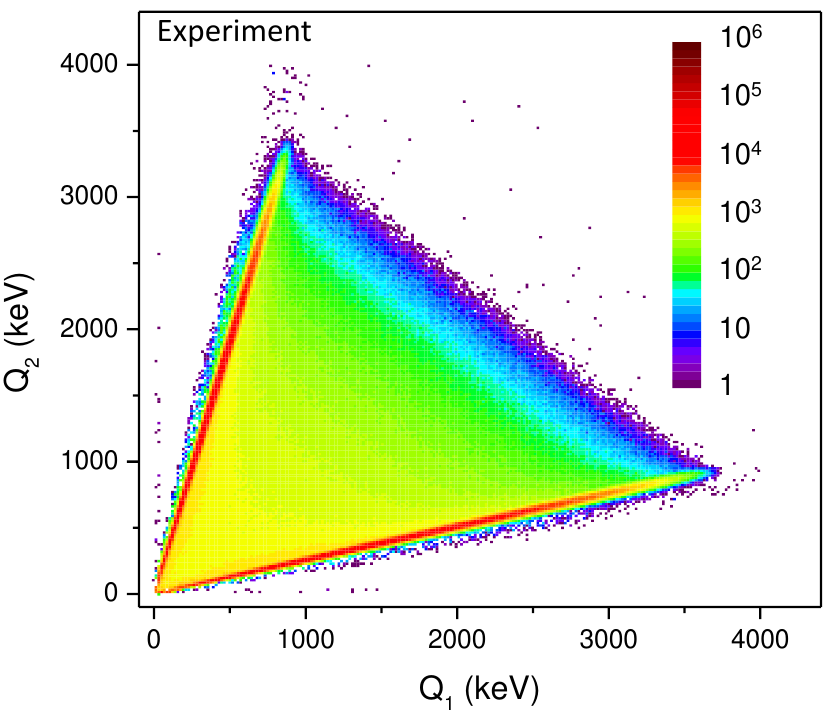}
\caption{\label{fig:Q1Q2exp}
2D histogram registered during the first half of the decay window showing
the energy collected in Det.~2 ($Q_2$) versus the
energy collected in
Det.~1 ($Q_1$), after background subtraction and a crude energy calibration.}
\end{figure}

For each event, the primary signals are the charges collected from the PMTs
along with the time stamp within the implantation-decay cycle.
A first crude calibration of the two YAP detectors was obtained using
the 59.54~keV peaks from the $^{241}$Am sources, without
involving any offset. As previously described in Ref.~\cite{Kan22}, this first calibration step includes a percent level rate-dependent correction of the PMT gains.  
This crude calibration is approximate at
the few percent level when applied to decay
electrons, mainly due to the difference in light collection efficiency
for scintillation
photons emitted near the $^{241}$Am source compared to those emitted
close to the front surface of the detector, where $^6$He is implanted.
Moreover, this
calibration does not account for light crosstalk and
cannot be applied to events where the total energy is shared between the
two detectors.
The resulting 2D energy histogram of the two collected energies is shown in
Fig.~\ref{fig:Q1Q2exp} for events registered during the first half of the
decay window (0 to 6~s), which is dominated by $^6$He decay events,
after subtraction of events in the second half of the window (6 to 12~s)
which is dominated by background.

\begin{figure}[!htb]
\includegraphics[width = 1.\columnwidth]{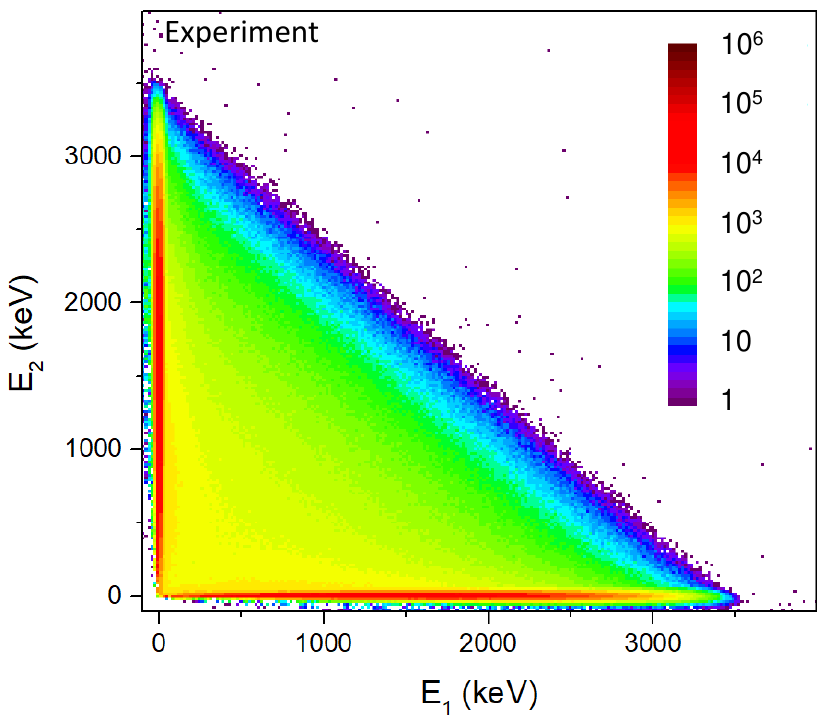}
\caption{\label{fig:E1E2exp}
2D histogram collected during the first half of the decay window, after
background subtraction, calibration and optical crosstalk correction.}
\end{figure}

The two most intense distributions
which look like lines in Fig.~\ref{fig:Q1Q2exp},
correspond to electrons
which deposited their full energy either in Det.~1 or in Det.~2.
The slopes of these lines
result from the light crosstalk, in which part of the light emitted by one
detector is
collected by the PMT of the other detector (Fig.~\ref{fig:Det}).
Events registered between these two lines are then attributed to electrons
that scattered from one detector to the other.
For these events, determining the energy deposited in each
scintillator requires the light transmission efficiencies to be known.
These are denoted by coefficients $f_{ij}$ (Fig.~\ref{fig:Det}),
where $i$ indicates the PMT that collected the light
and $j$ the YAP scintillator where the light was produced.
A refined calibration procedure has been applied to account for this crosstalk,
and is detailed in Appendix \ref{app:crosstalkAndResolution}. 
The resulting 2D histogram after background subtraction and calibration, including
the crosstalk correction, is shown in Fig.~\ref{fig:E1E2exp}.

The histogram in Fig.~\ref{fig:E1E2exp} provides
information on both, the nature of the event (backscattered or not) and the
total energy deposited, $E_{\rm sum} = E_1 + E_2$.
For a given total energy,
the fraction of events between the two main distributions provides then an
estimate of the backscattering probability.

\section{Geant4 simulations}
\label{G4sim}
\subsection{Geometry and event generator}
\label{sec:Geom}

Geant4 (v11.1.2) was used to generate $10^9$ events of beta particles
 following the decay spectrum of $^6$He,
 and to record the energies, $E_1$ and $E_2$, deposited in the YAP detectors.
 The particles were emitted isotropically, over a $4\pi$~sr solid angle,
 from a source located 130~nm inside
 Det.~1.
 The shape of the electron source was assumed to be a uniform disc, 
 with \O~4mm (determined by the last collimator), along the radial direction, and a 
 Gaussian distribution, with 47~nm standard deviation along the beam direction
 (from TRIM calculations~\cite{TRIM}).
 The energy spectrum of beta particles from $^6$He decay
 incorporated all corrections to the phase space, including radiative and hadronic
 ones \cite{Kan23}. The implemented setup geometry
 is shown in Fig.~\ref{fig:Det}, and
 includes the two YAP crystals, the PVT scintillators, the PMTs, and the aluminum
 detector housing.

\subsection{Physics Lists}
\label{sec:PhysLists}

The following
Physics Lists, included in the Geant4 package, have been considered in
the simulations, along with
the abbreviations used in the text below:

\begin{itemize}
\item G4EmStandardPhysics (Option0) 
\item G4EmStandardPhysics\_option3 (Option3) 
\item G4EmStandardPhysics\_option4 (Option4) 
\item G4EmLivermorePhysics (Livermore)
\item G4EmPenelopePhysics (Penelope)
\item G4EmStandardSS (SS)
\end{itemize}

Option0 is the default electromagnetic (EM) configuration optimized
for high-energy physics
and is expected to be applicable in processes down to 100~keV.
It is based on the Urban Multiple Scattering Model (UrbanMsc) \cite{Urb06}
for electrons and positrons.
Livermore and Penelope are specific configurations optimized for low-energy processes
and Option4 is a combination of the most accurate EM models. All of these use the
Goudsmit-Saunderson multiple scattering model for electrons and are
recommended for low-energy precision studies. Note that a new and more performant version of this model was fully implemented in version v10.6 of Geant4 \cite{Nov24}.
Option3 is also based on the UrbanMsc model, but is expected to be more accurate
than Option0 for the low-energy regime. It is faster than Option4 and may be
considered as an intermediate alternative between Option0 and Option4.
SS, based on a single scattering model, is by far more demanding
in process time, but is expected to yield the most accurate results for
electron scattering. For Option4, Penelope, and Livermore, a combination of multiple
scattering and single scattering is implemented for large-angle scattering.
Livermore and Option4 use the same models for electrons, and so provided
equivalent results in all aspects. Only the results from Option4 will thus
be shown in the following.

Default values were adopted for the relevant tracking parameters and for
the secondary particle creation energy thresholds.
For all but SS, 10$^9$ events were generated whereas for SS,
10$^8$ events were used because of the much longer computing time.
For illustration, the CPU time required for the simulation of 10$^6$ events
is reported in Table~\ref{tab:CPU} for each Physics List. Those obtained with
Option4, Livermore, and Penelope, which are
optimized for low-energy processes, are larger
by less than a factor of two compared to Option0 and Option3. With SS,
the process time is larger by more than two orders of magnitude.

\begin{table}[!htb]
    \caption{\label{tab:CPU} Process (CPU) times using Intel\textregistered Xeon\textregistered Gold 6246R CPU @ 3.4~GHz processors for the simulation of 10$^6$ $^6$He decay events, under the conditions described in Sec.~\ref{sec:Geom} for the six considered Physics Lists.}
    \centering
    \begin{tabular}{l@{\hspace{8mm}}c}
        \hline\hline
        EM Physics List & CPU time (s) \\ \hline
        Option0 & 1594 \\
        Option3 & 1636 \\
        Option4 & 2692 \\
        Livermore & 2608 \\
        Penelope & 2722 \\
        SS & 640653 \\ \hline\hline
    \end{tabular} 
\end{table}

\subsection{Calculated Backscattering Probabilities}
\label{sec:CalcBackProb}

Consider a source with electrons emitted over an hemisphere, with
solid angle $\Omega_1 = 2\pi$~sr. A common definition of the backscattering
probability, which is adopted here
in a first calculation, is given by the ratio between the
number of particles detected in the other
hemisphere relative to the total number of particles emitted
over $\Omega_1$.
To determine the backscattering probabilities from the simulations,
electrons generated over the $2\pi$ hemisphere covered by Det.~1,
were selected among all the isotropically generated events. This detector
is then considered as the ``target'' medium.
We define $\theta$ as the angle between the $^6$He beam direction and
the initial direction of the emitted beta particle. This
coincides with the emission angle
relative to the cylindrical axis of the YAP detector.

In the simulations, an event was tagged as backscattered when
some energy
was deposited in Det.~2. Such a definition includes naturally
events for which, at any step during slowing down, the
electron deposits some of its energy into Det.~2, even
when it finally stops in Det.~1. It also
includes events for which the electron does not actually
escape from Det.~1 but, energy from secondary particles
(bremsstrahlung photons, X-rays, etc.) is deposited in Det.~2.
Following the definition above, the backscattering probability
is then calculated from the ratio between the
number of events for which some energy was
deposited in Det.~2 relative to the number of generated events
over the $2\pi$~sr hemisphere.
This number can be considered within some
energy or angular window.
For later use, we denote by $R_{\rm sim1}$ these probabilities.

\begin{figure}[!htb]
\includegraphics[width = 1.\columnwidth]{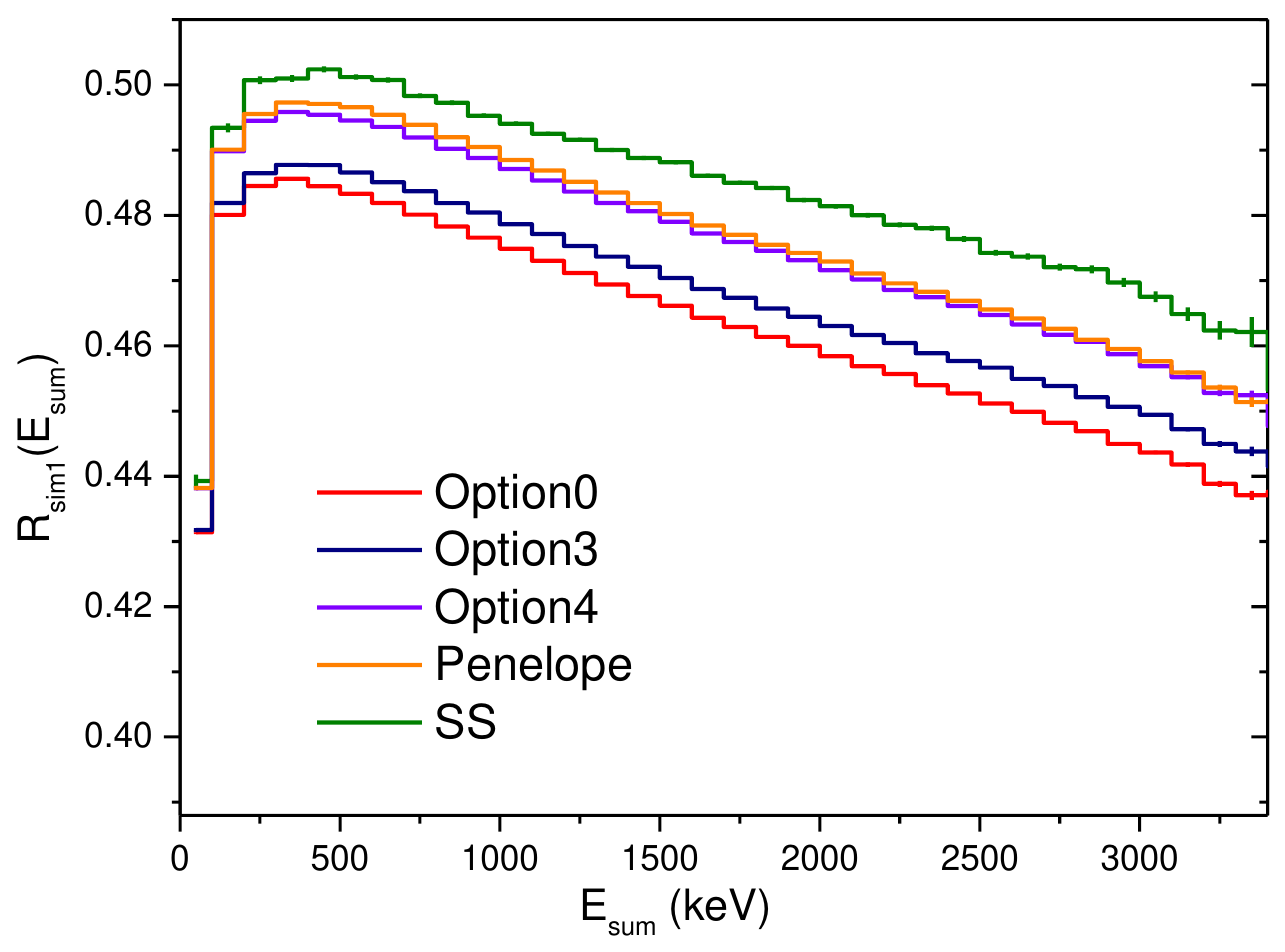}
\caption{\label{fig:probaEnergyG4}
Backscattering probability $R_{\rm sim1}$
obtained from Geant4 simulations as a function of the total deposited energy,
for five Physics Lists.
The energy bin width is 100 keV. The error bars are only statistical, at $1\sigma$,
and are visible at the extremes of the spectrum.}
\end{figure}

\begin{figure}[!hbt]
\includegraphics[width = 1.\columnwidth]{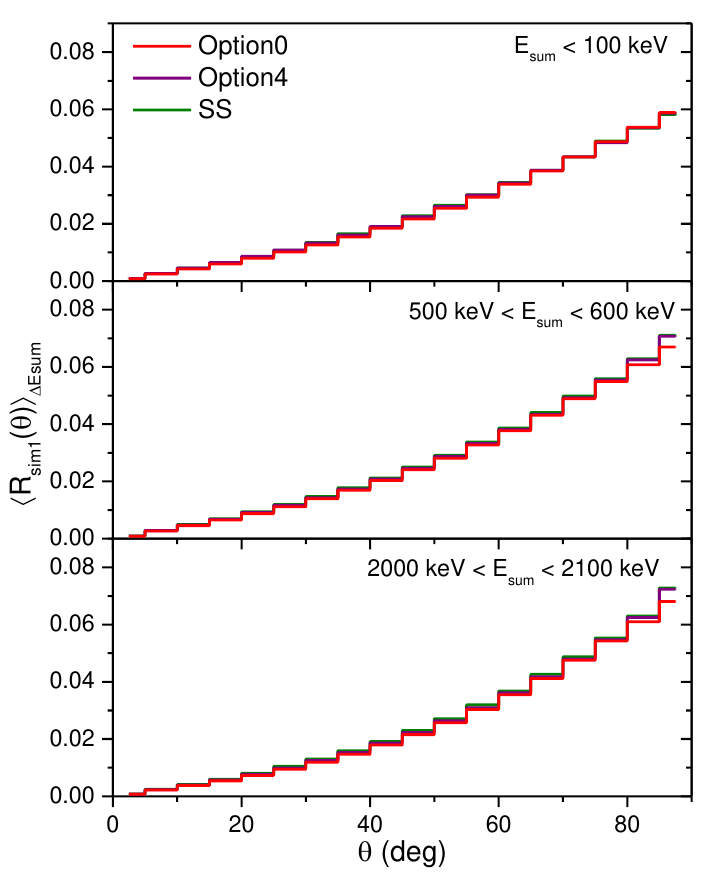}
\caption{\label{fig:probaAngleG4}
Backscattering probabilities $R_{sim1}$, weighted by the
solid angle as a function of incident angle $\theta$ for an isotropic
emission over $2\pi$~sr and for three energy intervals $\Delta E_{sum}$.
The bin width is $5^\circ$.}
\end{figure}

The probabilities $R_{\rm sim1}$ as a function of energy, obtained as
described above for the different Physics Lists,
are shown in Fig.~\ref{fig:probaEnergyG4} as a function of the
total deposited energy.
We observe that they range from 0.43 to 0.51, with a maximum
reached for an initial energy close to 500~keV.
It is notable that more refined and more computationally
demanding models predict larger backscattering probabilities. The relative
difference between the two extreme models (SS and Option0) is of the order
of 4\% to 5$\%$
between 500~keV and 3500~keV, and approximately $2\%$ for energies below 100~keV. 

The backscattering probabilities as a function of the initial
angle $\theta$,
weighted for the solid angle, are shown in Fig.~\ref{fig:probaAngleG4} for
three intervals of the total deposited energy.
Only results from Option0, Option4 and SS are shown since
the results from Penelope were found to be similar to those from Option4, whereas
Option3 provided intermediate results between those of Option0 and Option4.

As expected, the backscattering probability increases with incident angles.
Due to the larger solid angle with increasing $\theta$, the relative contribution
of large incidence angles is enhanced. 
At low energies (Fig.~\ref{fig:probaAngleG4}, top panel), the models do not display
any difference at this level of precision. Above 500~keV (Fig.~\ref{fig:probaAngleG4},
middle and bottom panels), the dominant deviation between Option0 and the other
models is observed at angles larger than $80^\circ$.
This highlights the significant impact of large incidence angles,
in particular, above $80^\circ$, whose contributions
are included in the present study.

\section{ANALYSIS METHOD}

The selection of events with $\theta \leq \pi/2$, made in the simulation,
cannot be applied to the experimental data for which the only available information
is the energy deposited in each detector.
In this section, we present the method used for the analysis of the experimental
spectrum shown in Fig.~\ref{fig:E1E2exp} to extract the backscattering probabilities.
The description of the method is done
using a simulated 2D histogram,
obtained with Option4 and shown in Fig.~\ref{fig:E1E2G4}, which includes the
convolution with the detector response, as described in Appendix \ref{app:crosstalkAndResolution}.

\subsection{General Principle}
\label{GeneralPrinciple}
\begin{figure}[!htb]
\includegraphics[width = 1.\columnwidth]{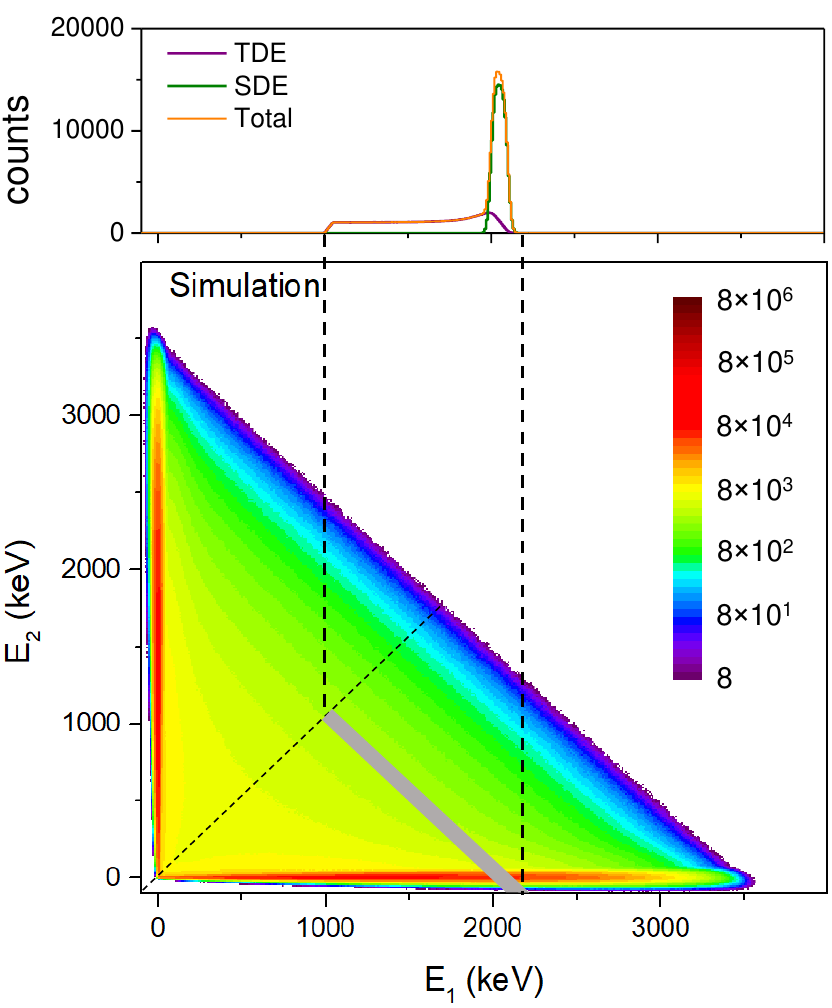}
\caption{\label{fig:E1E2G4}
2D histogram obtained from simulations with Option4, after convolution with the
detector response. The gray band indicates the selection of events with total deposited energy between 2000~keV and 2100~keV with $E_1 \geq E_2$. The upper
panel shows the distributions SDE (green), TDE (purple) and total (orange)
as a function of $E_1$.}
\end{figure}

Consider events with an energy deposited in Det.~1 which is larger
than the energy deposited in Det.~2. 
These events are located below the main diagonal line in the 2D histogram
of Fig.~\ref{fig:E1E2G4}.
The dominant horizontal distribution, which appears as a line,
is again due to electrons depositing their full energy in Det.~1, without backscattering.
These events will be termed ``single-detector events'' (SDE). 
Due to the very shallow implantation depth (130~nm) in the volume of Det.~1,
these events can, to first order, be assimilated to electrons initially emitted
towards Det.~1.
Since simulations give access to the exact deposited energy (before convolution
with the detector resolution), SDE are unambiguously identified in the simulated data by
imposing no energy deposition in Det.~2.
The other events, distributed over the triangle below the main diagonal
and the SDE, are labeled ``two-detector events''
(TDE) and arise from electrons backscattering in one of the two detectors. 
Simulations indicate that a significant fraction ($\approx 24\%$) of electrons
emitted towards Det.~1 that are backscattered in Det.~2 actually deposit more
energy in Det.~2 than in Det.~1.
However, because of the almost fully symmetric detection setup, those
events are
mirrored by electrons emitted towards Det.~2 and backscattering with a larger
energy deposited in Det.~1, and are then part of TDE.
In the following, we thus assimilate TDE with backscattered events for which
$E_1 \geq E_2$
and SDE with non-backscattered events. 

Next, we select a narrow energy interval around a given total
energy, $E_{\rm sum} = E_1 + E_2$ (grey band in Fig.~\ref{fig:E1E2G4}). Since the
simulation provides the identification of the event,
it enables to build separately
the distributions for SDE and TDE (upper panel in Fig.~\ref{fig:E1E2G4}).
If $N_{\rm sim}^{\rm SD}$ and $N_{\rm sim}^{\rm TD}$ denote respectively the
number of SDE and TDE
obtained from the simulations, then the ratio
\begin{equation}
\label{eq:Rsim2}
    R_{\rm sim2} = \frac{N_{\rm sim}^{\rm TD}}{N_{\rm sim}^{\rm SD} + N_{\rm sim}^{\rm TD}}~\,,
\end{equation}
provides an estimate of the backscattering probability.
We note that
this ratio is independent of the convolution since the simulation gives
unambiguously access to the type
of event, SDE or TDE, before applying the convolution with the detector response
function. In other words, $N_{\rm sim}^{\rm TD}$ and $N_{\rm sim}^{\rm SD}$, are the
integrals
of the distributions shown in the upper panel of Fig.~\ref{fig:E1E2G4}.

The procedure above is then repeated for each Physics List to build the simulated 2D histogram from which the SDE and TDE distributions are determined.

It is worth noting that the 130~nm implantation depth of the $^6$He ions inside Det.~1, implemented in the simulation, breaks the symmetry of the detection system. The ratio $R_{\rm sim2}$, which is directly comparable to experimental data, is therefore not strictly equivalent to the ratio $R_{\rm sim1}$ which quantifies the real backscattering probability of electrons hitting the surface of a YAP crystal. This difference is discussed in
Sec.~\ref{sec:comparison}.

A similar analysis for events with $E_2 \geq E_1$ would have been valuable as well. However, a clean definition of a SDE for Det.~2 is impossible for such a condition. Due to the implantation in Det.~1, all electrons do deposit some energy in Det.~1 prior to reaching Det.~2. Such analysis would have required the use of an arbitrary energy threshold whose effect is difficult to interpret.

\subsection{Comparison Test}
\label{sec:comparison}

\begin{figure}[!htb]
\includegraphics[width = 1.\columnwidth]{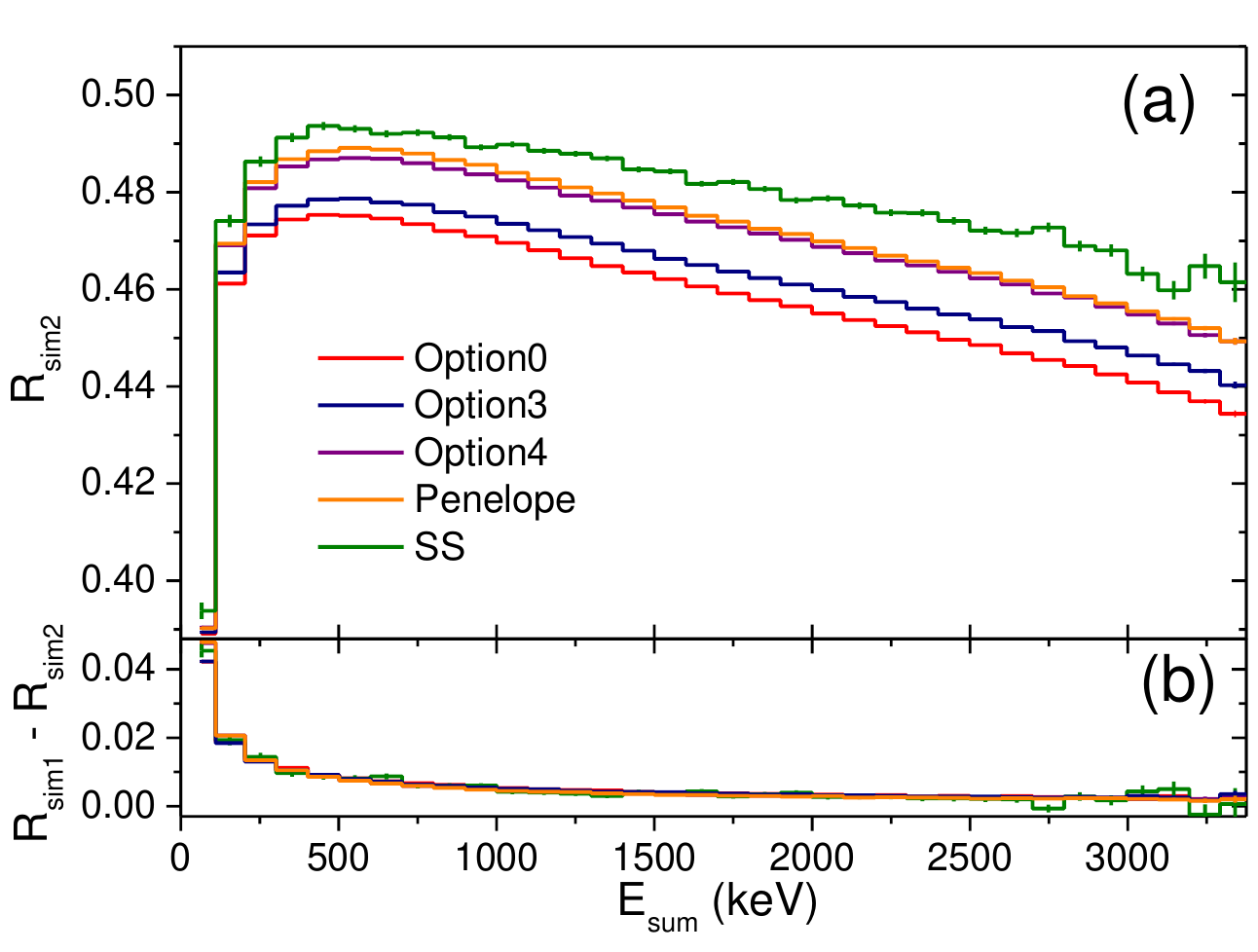}
\caption{\label{fig:RS2}
(a): Ratio $R_{\rm sim2}$ as a function of the total deposited energy
for the different Physics Lists; (b): Difference $(R_{\rm sim1} - R_{\rm sim2})$
between the backscattering probabilities deduced from Fig.~\ref{fig:probaEnergyG4} and from panel (a) above. The energy bin width is~100 keV. Error bars are only statistical, at $1\sigma$.}
\end{figure}

The ratio $R_{\rm sim2}$ is shown in Fig.~\ref{fig:RS2} (a) as a function of the
total deposited energy obtained with the five Physics Lists. 
This plot shows very similar features to those of Fig.~\ref{fig:probaEnergyG4},
where the $R_{\rm sim1}$ backscattering probabilities were determined by selecting electrons
emitted only towards Det.~1 and requesting energy deposition in Det.~2.
The ratio $R_{\rm sim2}$ is systematically lower than $R_{\rm sim1}$,
as shown in Fig.~\ref{fig:RS2} (b).
For the lowest-energy bin, centered at 50~keV, the absolute difference is
$\approx 0.04$. The difference quickly decreases to $\approx 0.008$ at 500~keV and further
down to $\approx 0.002$ at 3000~keV, for all Physics Lists.

All Physics Lists produce the same differences
$(R_{\rm sim1} - R_{\rm sim2})$
as a function of the total energy [Fig.~\ref{fig:RS2} (b)]. This suggests that
these discrepancies are related to the difference in the definition
of backscattering probabilities associated with $R_{\rm sim1}$ (Sec.~\ref{sec:CalcBackProb})
and $R_{\rm sim2}$ in Eq.~(\ref{eq:Rsim2}).
It also shows that both definitions are equally
valid for testing and comparing the different Physics Lists.
This difference is due to the small but still significant implantation depth
of $^6$He in Det.~1. 
A small fraction of the electrons emitted towards Det.~2 can backscatter
before reaching Det.~2 and leads to a larger value of $N_{\rm sim}^{\rm SD}$ and hence
to a reduction of $R_{\rm sim2}$.

\subsection{Application to Experimental Data}

The same energy cuts as those applied to the simulation histograms,
are then applied to the experimental 2D histogram in Fig.~\ref{fig:E1E2exp}.
The projection of
events along the $E_1$ axis gives a single distribution which contains an unknown mixture of SDE and TDE events. To unravel the two contributions, this single distribution is
then fitted with a linear combination of the SDE and TDE
distributions deduced from the simulation, with the same energy cuts
and accounting for the detector resolution as determined in Appendix B.
From the fitted distributions, one deduces then the integrals
$N_{\rm exp}^{SD}$ and $N_{\rm exp}^{TD}$ which are used to calculate the
corresponding experimental ratio
\begin{equation}
R_{\rm exp} = \frac{N_{\rm exp}^{\rm TD}}{N_{\rm exp}^{\rm SD} + N_{\rm exp}^{\rm TD}}~\,.
\label{eq:Rexp}
\end{equation}

\section{Results}

\begin{figure}[!htb]
\includegraphics[width = 1.\columnwidth]{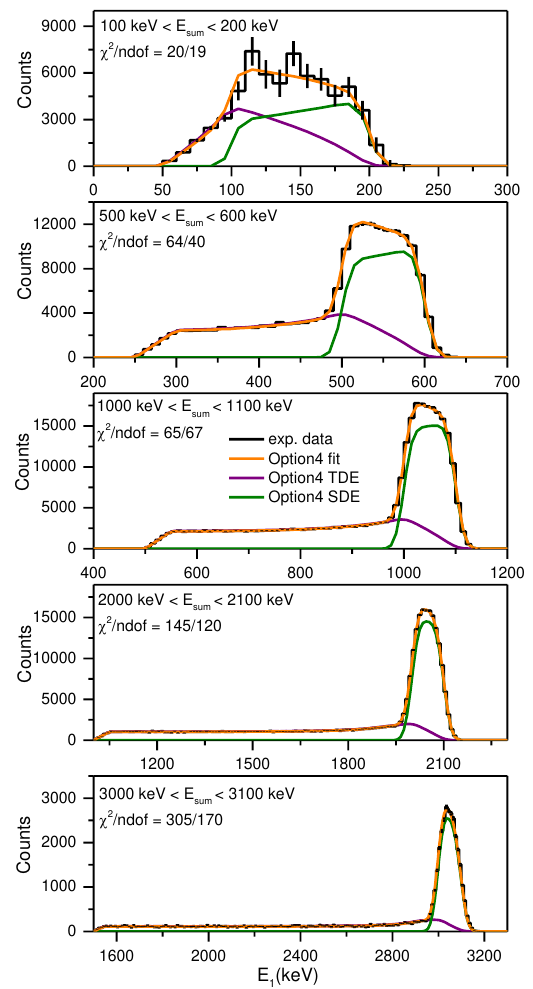}
\caption{\label{fig:E1_ExpOpt4}
Fit of the experimental data (black histograms) with
energy distributions obtained with Option4
for five intervals of the total deposited energy. The green and purple lines
correspond respectively to the SDE and TDE distributions.
Orange curves are fits to the data. The statistical uncertainties
are at $1\sigma$.}
\end{figure}

The energy distributions projected along $E_1$, for the
simulation and the experimental data, have been analyzed for 33 windows
of the total deposited energy. The lower edge of the windows ranged
from 100 to 3300~keV and
the window width was 100~keV.
Data with a total energy smaller than 100~keV 
or larger than 3400~keV were discarded due to poor statistics.
The distributions obtained using Option4 are compared to the experimental ones in Fig.~\ref{fig:E1_ExpOpt4} for five energy windows. 
The lines in green and purple correspond, respectively, to SDE and TDE from the simulation.
The histograms in black show the experimental data, after normalization to
the total number of simulated events.
The orange lines represent the fits of the experimental spectra using
a linear combination of SDE and TDE distributions from simulations.

We denote by $s_1$ and $s_2$ the scaling factors for, respectively, SDE
and TDE distributions. These are the only free parameters of the fits to
precisely adjust
both the backscattering probability and the number of events in each energy selection.
The experimental ratio in Eq.~(\ref{eq:Rexp}) has then been estimated using
the fitted parameters, with

 \begin{eqnarray}
 N_{\rm exp}^{\rm SD} &= s_1 N_{\rm sim}^{\rm SD} \label{eq:NeSD}~\,, \\
 N_{\rm exp}^{\rm TD} &= s_2 N_{\rm sim}^{\rm TD} \label{eq:NeTD}~\,.
 \end{eqnarray} 

In Fig.~\ref{fig:E1_ExpOpt4}, the backscattering tail (purple line) from TDE
nicely follows the experimental data for all five energy windows.
The result of the fit (in orange) also matches the experimental data
for all windows and over the
full spectra. This demonstrates not only that the
simulation using Option4 reproduces remarkably well the relative amount of
backscattered
events but also the energy distribution between the two detectors.

\begin{figure}[!htb]
\includegraphics[width = 1.\columnwidth]{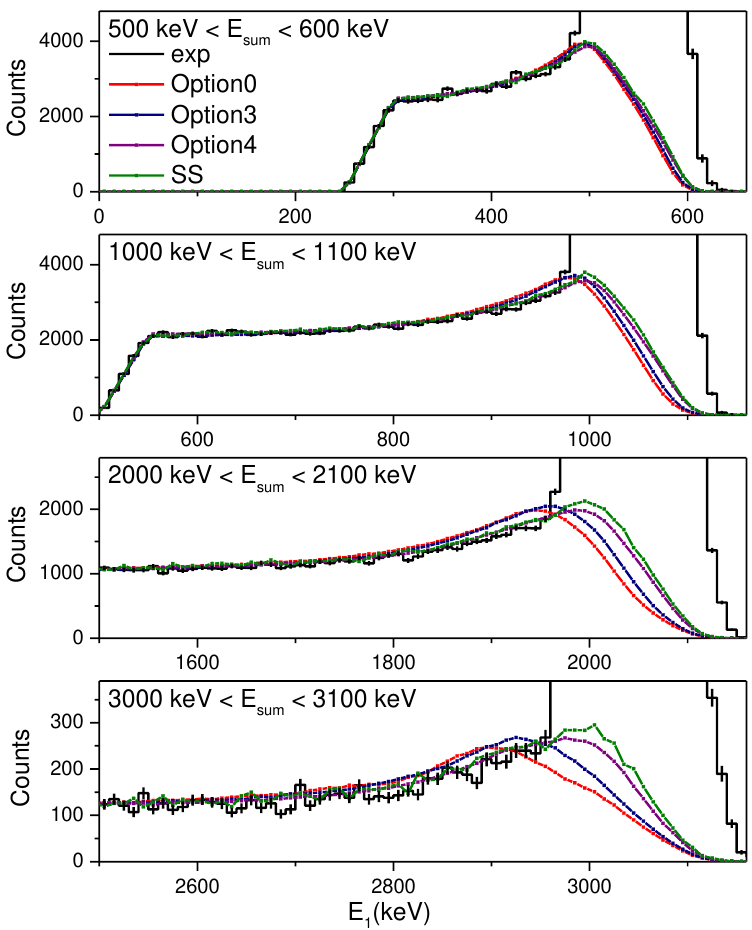}
\caption{\label{fig:E1_TailExpModels}
Comparison of the TDE distributions obtained from simulations,
using the different Physics Lists, with experimental data normalized to
the total number of simulated events. The uncertainties on the experimental
spectra are at $1\sigma$.}
\end{figure}

The tails corresponding to TDE obtained with four different Physics Lists
are compared to the experimental data in Fig.~\ref{fig:E1_TailExpModels} for
four windows of the total deposited energy.
The distributions obtained with Penelope were equivalent to those obtained
with Option4 and are not shown.
For all energy windows, the results obtained with Option4 and SS reproduce
the experimental data fairly well.
This is not so for Option0 and Option3, which overestimate the backscattering
tail in the region of the spectrum at 100~keV to 200~keV below the main
SDE peak. The discrepancies also increase at higher total energies.
At the same time, simulations with Option0 and Option3 underestimate
backscattering events, depositing less than 50~keV in Det.~2,
located in the region underneath the SDE peak.

Finally, the experimental backscattering probability, $R_{\rm exp}$,
was extracted using the distributions obtained from each simulation and the
fitting procedure described above.
The ratio $R_{\rm sim2}$ was then compared
to the corresponding $R_{\rm exp}$ ratio extracted, in a self-consistent way, using
the distributions from that same simulation. Note that the shape differences for
the TDE distributions shown in Fig.~\ref{fig:E1_TailExpModels} also affect the ratio
$R_{\rm exp}$ extracted using the different models.

\begin{figure}[!htb]
\includegraphics[width = 1.\columnwidth]{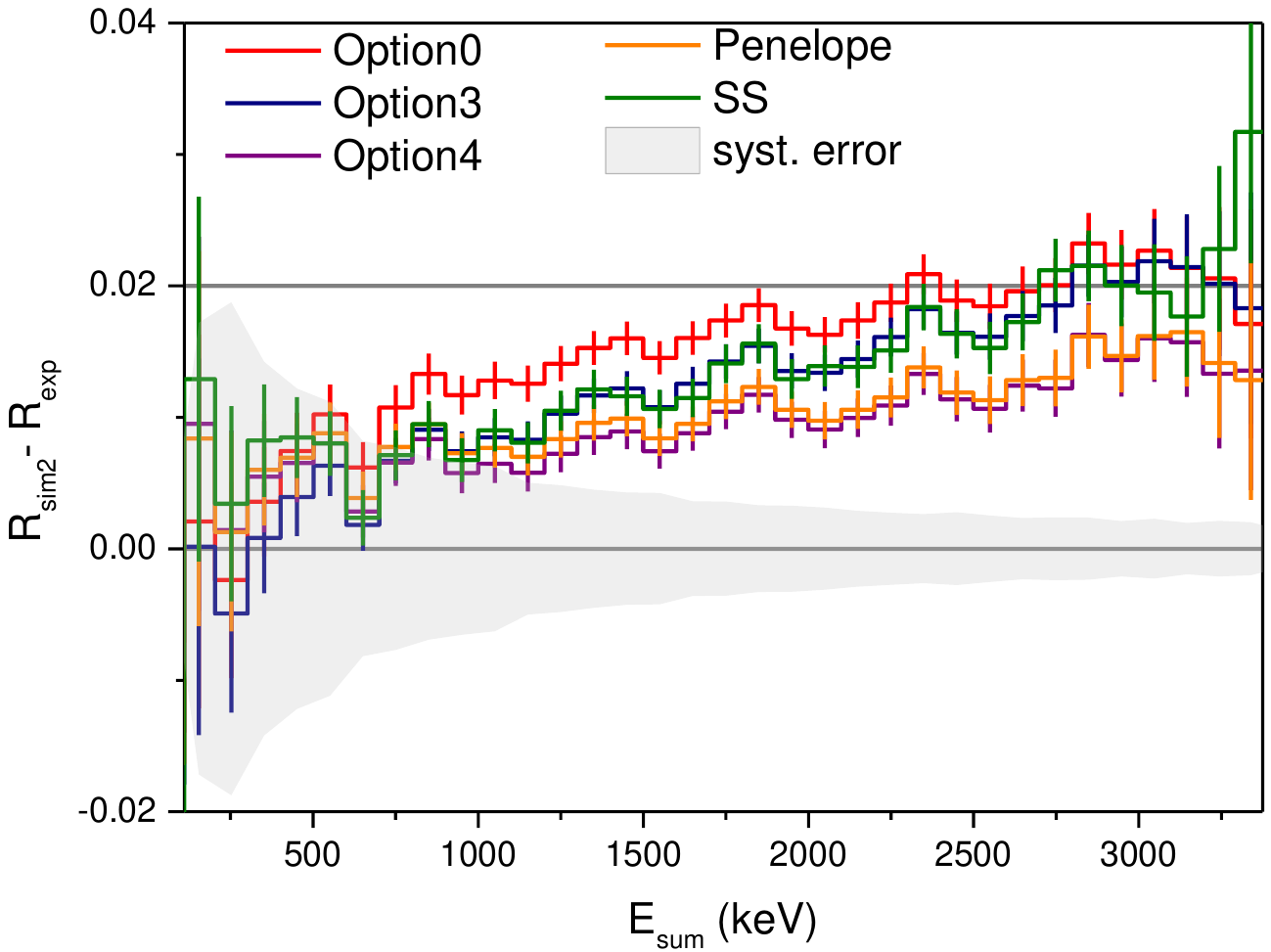}
\caption{\label{fig:RDiff}
Difference between the ratios $R_{\rm sim2}$ and $R_{\rm exp}$ obtained using the different Physics Lists. The statistical uncertainties are at $1\sigma$.
See text for details.}
\end{figure}

Figure \ref{fig:RDiff} shows the difference, $(R_{\rm sim2} - R_{\rm exp})$ between the ratios
from simulations and from experimental data for each Physics List. The
error bars indicate the statistical uncertainty, dominated by the experimental data.
The light gray area indicates the systematic uncertainty due to a possible calibration
mismatch of at most 2.5~keV.

For a total deposited energy below 750~keV, all simulations are consistent with
the experimental data, although the statistical uncertainties are of the order of $1\%$.
Above 750~keV, the ratio $R_{\rm sim2}$ is systematically higher than $R_{\rm exp}$ for all
Physics Lists.
This indicates an overestimate of the backscattering probability by the simulations.
Nevertheless, the deviation from the experimental value does not exceed $2\%$ for the largest
(Option0), which corresponds to a relative overestimate by less than $5\%$. For Option4 and
Penelope, the deviations are even smaller, with a relative overestimate of the backscattering probability remaining below $3.5\%$. We remind here that the Livermore Physics List provided results equivalent to those of Option4. 
Quite surprisingly, the simulation using SS, which was anticipated to be the most
accurate, does not outperform Option 4 and Penelope in terms of agreement with experimental data.

\section{CONCLUSIONS}

We have determined the backscattering probabilities of electrons impinging
on a YAP scintillator with energies from 0.1 to 3.4~MeV.
The measurement used
a $2\times2\pi$ calorimeter, which was originally designed to eliminate
the effect of electron backscattering in measurements of beta-energy spectra.
The separation of the calorimeter into two halves enabled the precise
identification of backscattered events.

The quantitative analysis involves comparisons with simulations to benchmark six
Physics Lists incorporated in Geant4.
For Option4, Livermore and Penelope, simulations are closer to experimental data
in the explored energy range,
with relative deviations on the backscattering probabilities reaching $3.5\%$
for the largest incident energies. 
Moreover, the distributions of the energy loss for backscattered events
were properly reproduced with
these three Physics Lists as well as with SS. 
For Option0 and Option3, which are not optimized for low energy processes, the
agreement remains quite acceptable, with relative differences in backscattering
probabilities below $5\%$. However, the shape of the experimental backscattering
tail is not properly reproduced by these last two Physics Lists.
The backscattering probabilities obtained with SS were surprisingly in poorer
agreement with experimental data than those with Option4, Livermore and Penelope.

The uncertainties on electron backscattering often represent a significant part
of the systematic uncertainty budget for precision measurements in beta decay. Due to the lack of
precise benchmarks of Geant4 for the backscattering of electrons in the energy range
from 100~keV up to several MeV, conservative relative uncertainties of the order of $10\%$
to $20\%$ have usually been considered. 
This is particularly the situation for experiments using strong magnetic fields, in which
the detection of beta particles produces large incident angles \cite{Ara20}.

The agreement obtained in the present work between simulations and
experimental data,
which covers a large energy range and includes large incident angles, is rather remarkable.
This invites a revision of the conservative approach
for estimating the impact of
simulations, especially when using the most accurate options such as Option4, Penelope or
Livermore. Although the results presented here cannot readily be transposed to other
materials or to any arbitrary geometry, it should help mitigating the usually large relative
uncertainties considered for backscattering probabilities.

\section{Acknowledgements}

The authors thank the technical support from LPC Caen and GANIL staff and
are grateful to A. Singh for her assistance during the running of the experiment.
This project was supported in part by the Agence Nationale de la Recherche
under grant ANR-20-CE31-0007-01 (bSTILED). 

\section{Disclaimer}
Certain trade names and company products are mentioned in the text or identified
in illustrations in order to adequately specify the experimental procedure and
equipment used. In no case does such identification imply recommendation or
endorsement by the National Institute of Standards and Technology nor does
it imply that the products are necessarily the best available for the purpose.

\appendix
\section{Background subtraction}
\label{app:bckgd_sub}

\begin{figure}[!htb]
\includegraphics[width = 1.\columnwidth]{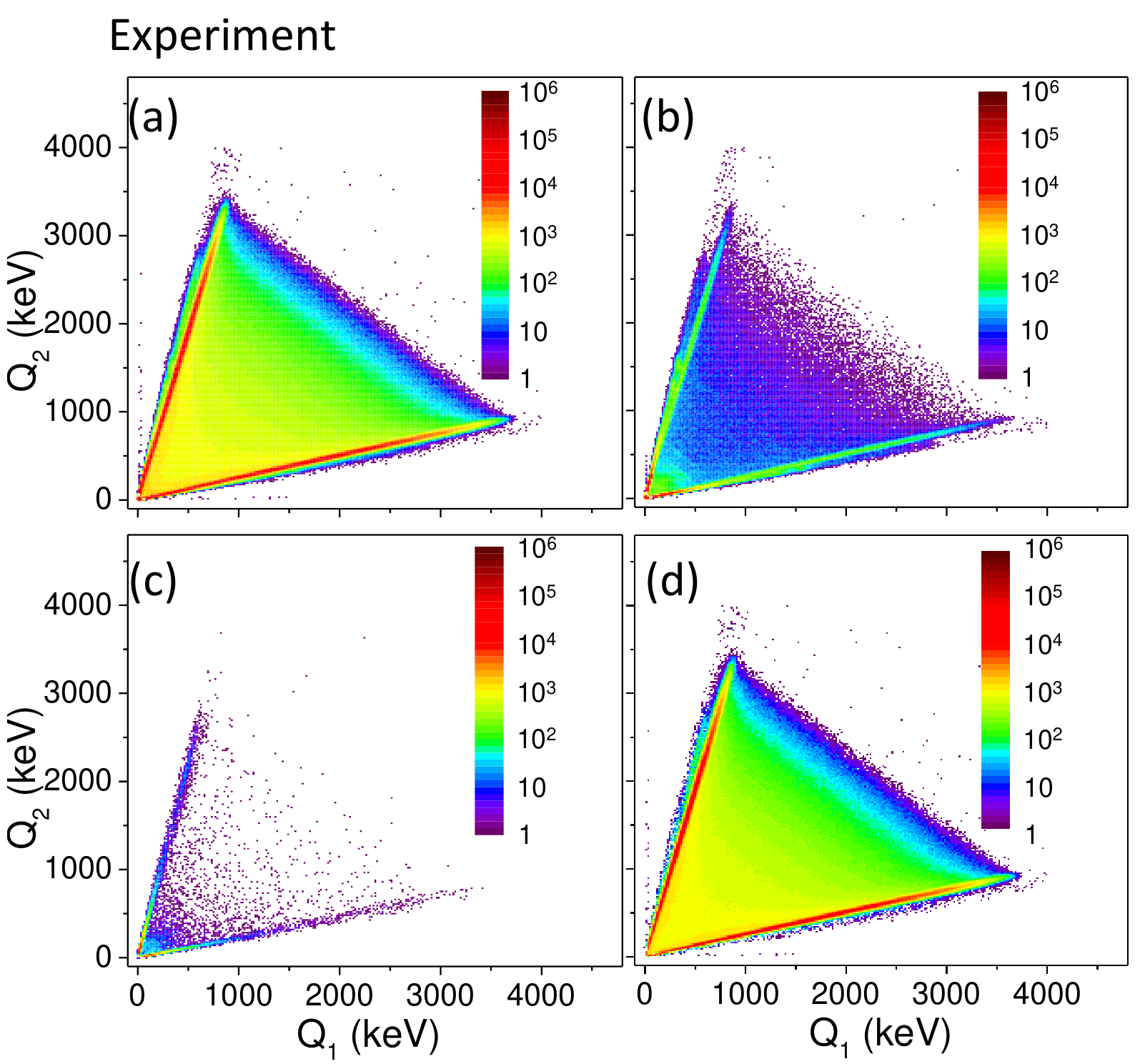}
\caption{\label{fig:Q1Q2exp-4plots}
Experimental 2D histograms for different conditions of the experiment.
(a): data from the first half of the decay window (0 to 6~s) dominated
by electrons from $^6$He decay; (b): data from 
the second half of the decay window (6 to 12~s) dominated by the constant
background; (c): data from a dedicated run performed to identify background sources
in the absence of implantation; (d): data of the first half of
the decay window after background subtraction.}
\end{figure}

Previous analyses of the $^6$He decay time spectra performed with the present
setup \cite{Kan22}, enabled the identification of two
main background contributions. The first was due to the constant ambient background,
which includes the contribution of the 59.54~keV $\gamma$ rays from $^{241}$Am. 
This background was suppressed by subtracting events in the
second half (6 to 12~s) of the decay window [Fig.~\ref{fig:Q1Q2exp-4plots} (b)],
from those in the first half (0 to 6~s)
of the decay window [Fig.~\ref{fig:Q1Q2exp-4plots} (a)]. 
The second background contribution arises from bremsstrahlung radiation produced
by decays of $^6$He nuclei implanted in the last collimator (Fig.1).
This contribution was subtracted using a dedicated background run, in which a 0.4~mm 
thick Al plate was placed at the exit of the collimator to prevent direct implantation
on Det.~1. 

The energy distribution of these bremsstrahlung events in the 2D
energy
histogram is shown in Fig.\ref{fig:Q1Q2exp-4plots} (c).
The normalization factor for this background subtraction was determined by suppressing its dominant contribution, peaked close to 100~keV. It was first determined with a few percents precision using the 2D spectra and was subsequently refined through a fit of the total energy spectrum after the refined calibration procedure, as described below.

\section{Crosstalk correction and detector resolution}
\label{app:crosstalkAndResolution}

Because of the optical coupling between the two detectors, their calibrations
and the determination of their response functions
have to be performed simultaneously.
After background subtraction, the experimental 2D histogram in Fig.~\ref{fig:Q1Q2exp-4plots} (d) was used, together with
Geant4 simulations, to determine the calibration parameters required
for an accurate reconstruction of the energies $E_1$ and $E_2$
deposited in each detector.
These parameters, noted $F_{ij}$ ($i,j=1,2$), are proportional to the relative
light collection coefficients $f_{ij}$ ($i,j=1,2$) illustrated in
Fig.~\ref{fig:Det}, and two offsets $Q_{\rm off1}$ and $Q_{\rm off2}$.
The light transport within the crystals was not included in the
simulations since
the experimental spectra provide the average transport coefficients.
The strong proportionality of YAP crystals compared to other
types of scintillators \cite{Mos98, Men98} enables the use of a linear relationship
between light production and deposited energy such that
\begin{equation}
\begin{pmatrix}
Q_1 \\ Q_2
\end{pmatrix}
=
\begin{pmatrix}
F_{11} & F_{12}\\
F_{21} & F_{22}
\end{pmatrix}
\begin{pmatrix}
E_1 \\ E_2
\end{pmatrix}
+
\begin{pmatrix}
Q_{\rm off1} \\ Q_{\rm off2}
\end{pmatrix}~\,.
\label{eq:Q1Q2}
\end{equation}

The Option4 simulation (Sec.~\ref{G4sim}~) was used for the refined calibration, with $10^9$ events. 
Estimates of the $F_{ij}$ coefficients were first inferred from the 2D
histogram in Fig.~\ref{fig:Q1Q2exp-4plots} (d) using the end point energies
and the slopes of the two main distributions,
corresponding to a full energy deposition either in Det.~1 or Det.~2.
In this first step, the offsets were considered negligible
and set to zero. Using these values, Eq.~(\ref{eq:Q1Q2}) was
then applied to the simulated events,
to compute the quantities $Q_1$ and $Q_2$.

The effect of the detector resolution was implemented in the
simulation by convolving $Q_1$ and $Q_2$
with Gaussian functions having standard deviations
$\sigma_1(Q_1)$ and $\sigma_2(Q_2)$. Considering that the dominant
contribution to the resolution is the statistical fluctuations associated
with the number of photoelectrons collected by the PMTs we have, for each detector,
\begin{equation}
\sigma_i(Q_i) = \alpha_i \sqrt{Q_i}~\,.
\label{eq:sigma1}
\end{equation}
In a first crude step, the values of $\alpha_i$ were
inferred from the fits of the 59.54~keV gamma peaks \cite{Kan22}.

\begin{figure}[!htb]
\includegraphics[width = 1.\columnwidth]{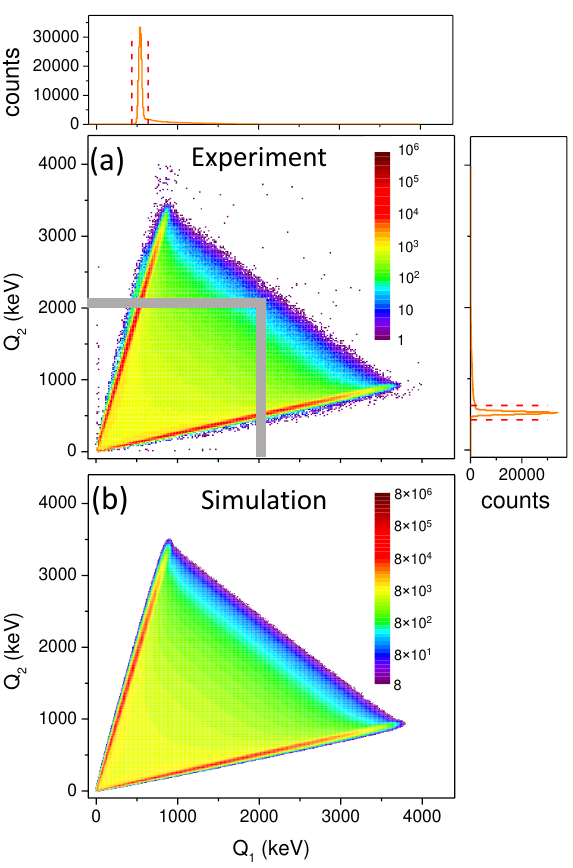}
\caption{\label{fig:Q1vsQ2}
(a): Illustration of the data selections used for the refined calibration.
The dashed red lines on the projected histograms indicate the range for
the Gaussian fits of the SDE distributions; (b): 2D energy histogram generated
with simulated events using the refined calibration parameters which account for
the detector resolution and crosstalk correction.
}
\end{figure}

This first procedure produced then a 2D histogram from the
simulated data [Fig.~\ref{fig:Q1vsQ2} (b)] closely matching
the experimental one [Fig.~\ref{fig:Q1vsQ2} (a)].
The refined calibration to improve the precision
on the $F_{ij}$ parameters and on possible offsets consisted in an iterative
procedure, which makes successive comparisons of the simulated and experimental histograms.
We defined 35 successive slices in $Q_1$ and $Q_2$, for the
experimental data, having 100~keV width,
as illustrated in Fig.\ref{fig:Q1vsQ2} (a).
Their projections are dominated by the SDE (see Sec.~\ref{GeneralPrinciple}), which were fitted with Gaussian
functions whose mean values, standard deviations and integrals were compared
to those obtained with simulated events. 
The fit range, indicated by dashed red lines in the projections
of Fig.\ref{fig:Q1vsQ2} (a), selects mostly SDE, so that the backscattering tail
only has a marginal effect in the calibration. The $F_{ij}$ values and
offsets were then updated after several iterations, to minimize the deviation
between the fit results from the experimental and simulated spectra.
The iteration process was stopped when the deviations on the mean values
remained below 2.5~keV over the full 3500~keV range and became dominated by
non-linear effects that are not yet implemented in the calibration functions. 
The $\alpha_i$ parameters were also adjusted during the iteration process,
resulting in a change of less than 3$\%$ from their initial values.
The final precision on the calibration was considered to be more than sufficient
for the present study of backscattering probabilities. The calibration parameters
obtained are listed in Table \ref{tab:fQalphas}.

\begin{table}[!htb]
    \centering
    \caption{Values of the calibration coefficients, offsets, and convolution
    parameters obtained from the iterative procedure of the refined calibration.} 
    \begin{tabular}{l@{\hspace{15mm}}l}
        \hline\hline
        $F_{11}$ = 1.0652   &   $Q_{\rm off1}$ = 2.0 keV \\
        $F_{12}$ = 0.2576   &   $Q_{\rm off2}$ = 3.5 keV \\
        $F_{22}$ = 0.9847   &   $\alpha_1$ = 0.61 keV$^{1/2}$\\
        $F_{21}$ = 0.2694   &   $\alpha_2$ = 0.65 keV$^{1/2}$\\
        \hline\hline
    \end{tabular}
    \label{tab:fQalphas}
\end{table}

\section{Total Energy Reconstruction}

The procedure described above consisted to match, as close as possible,
the simulated 2D histogram to the experimental one.
However, the specific shape of the $\beta$-particle energy spectrum is not
involved in the extraction of the backscattering probabilities. It is included
in the simulations in order to match the properties of the electron source.

\begin{figure}[!htb]
\includegraphics[width = 1.\columnwidth]{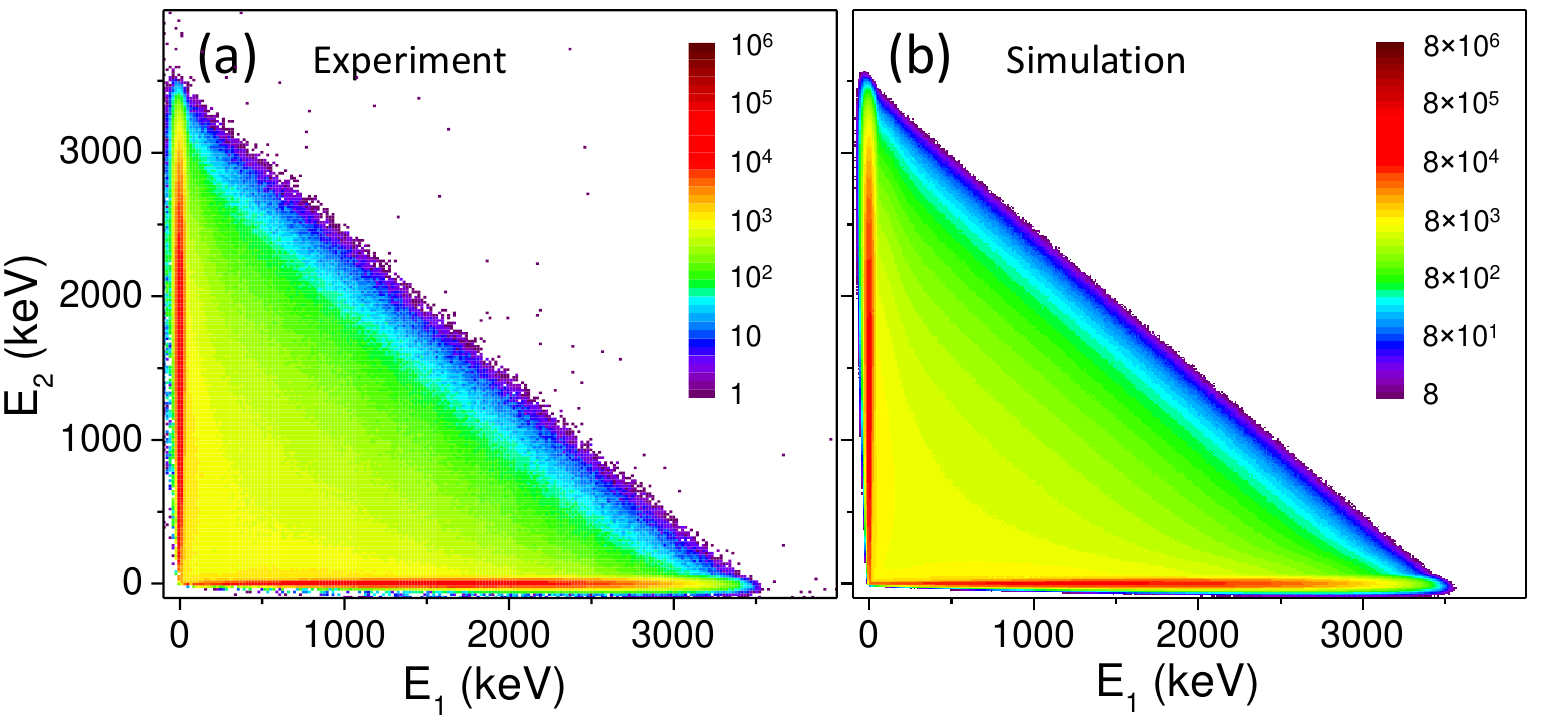}
\caption{\label{fig:E1E2_Comp}
Experimental (a) and simulated (b) 2D energy histograms of the energy deposited in Det.~1 and Det.~2. The experimental 2D histogram was obtained after background subtraction and includes
the refined calibration, whereas the simulated 2D histogram includes the convolution.}
\end{figure}

Using the set of parameters determined above and inverting Eq.~(\ref{eq:Q1Q2}),
the energies $E_1$ and $E_2$ deposited in each detector can be determined,
%
\begin{equation}
\begin{pmatrix}
E_1 \\ E_2
\end{pmatrix}
=
\frac{1}{d}
\begin{pmatrix}
F_{22} & -F_{21}\\
-F_{12} & F_{11}
\end{pmatrix}
\begin{pmatrix}
\Delta Q_1 \\ \Delta Q_2
\end{pmatrix}~\,,
\label{eq:E1E2}
\end{equation}
where $d = (F_{11}F_{22} - F_{12}F_{21})$ and $\Delta Q_i = (Q_i - Q_{{\rm off}i})$.
The total deposited energy, $E_{\rm sum} = E_1 +E_2$, can then be reconstructed.
The resulting 2D histograms obtained for both experimental and simulated data are displayed
again in Fig.~\ref{fig:E1E2_Comp}. These are the same 2D
histograms as
those in Figs.~\ref{fig:E1E2exp} and \ref{fig:E1E2G4}.

\begin{figure}[!htb]
\includegraphics[width = 1.\columnwidth]{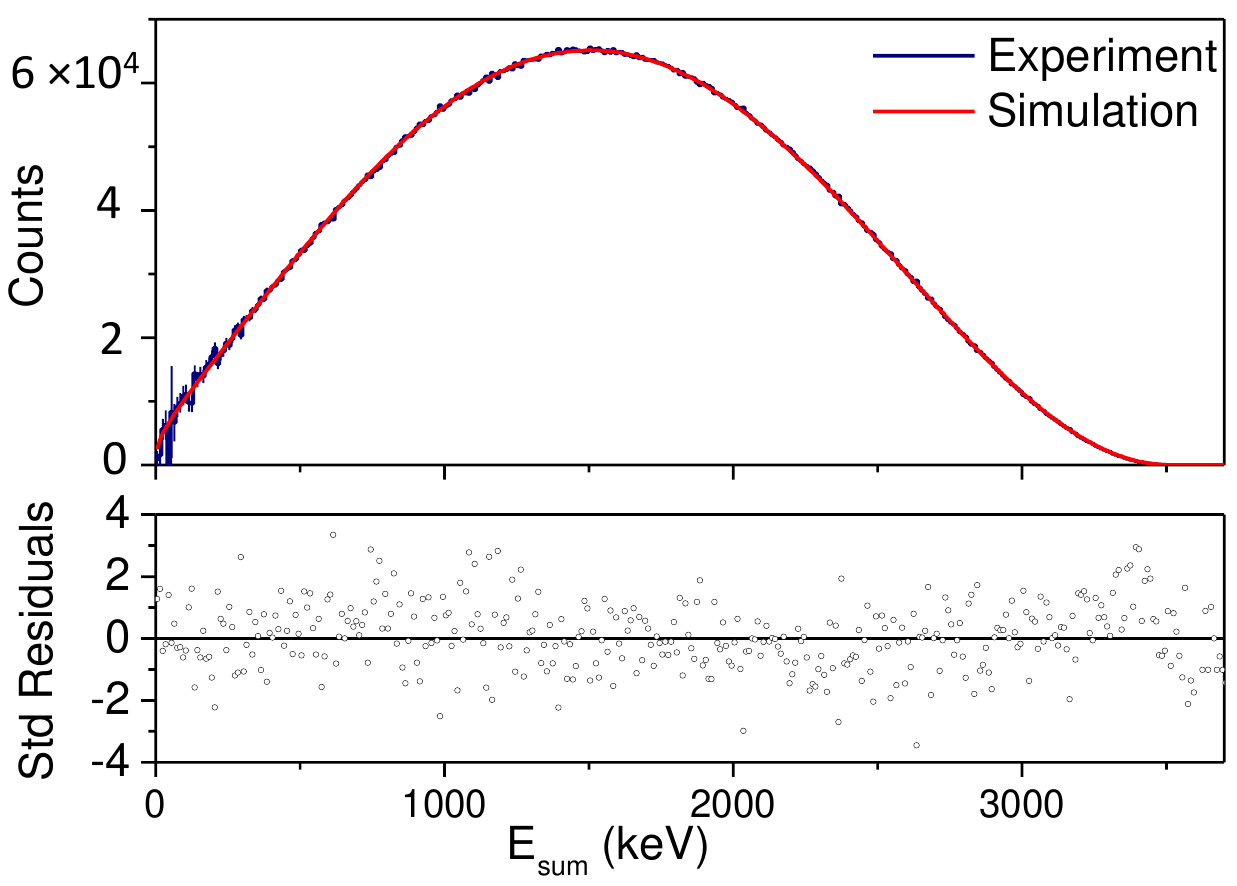}
\caption{\label{fig:Espectrum}
Upper panel: reconstructed total energy spectrum for the
experimental (blue) and simulated (red) data.
The larger statistical fluctuations below 300~keV are due
to the impact of the large bremsstrahlung background contribution.
Lower panel: standard residuals from the fit.}
\end{figure}

The experimental and simulated spectra of the total energy deposited are compared
in the upper panel of Fig.~\ref{fig:Espectrum}. The simulated spectrum was adjusted
to the experimental one using 
only two free parameters: the overall scaling factor and the normalization factor for
the subtraction of the bremsstrahlung induced background discussed in
Appendix \ref{app:bckgd_sub}.
The lower panel of Fig.~\ref{fig:Espectrum} shows the standard residuals from the fit.
The agreement
between the two spectra provides an additional internal
consistency check and validates both the background subtraction
and the calibration procedure. 

\newpage
\bibliography{backscattering}
\bibliographystyle{unsrturl}
\end{document}